\documentclass[proof]{WileyASNA-v1}

\articletype{Article Type}%

\received{26 April 2016}
\revised{6 June 2016}
\accepted{6 June 2016}

\begin{document}

\title{Looking for fast optical bursts from FRB121102:
case study for a small telescopes with sub-second temporal resolution}

\author[1,2,3]{S. Karpov*}

\author[4]{M. Jelinek}

\author[4]{J. \v{S}trobl}

\authormark{S.~Karpov \textsc{et al}}

\address[1]{\orgdiv{CEICO}, \orgname{Institute of Physics, Czech Academy of Sciences}, \orgaddress{\state{Prague}, \country{Czech Republic}}}

\address[2]{\orgdiv{Special Astrophysical Observatory}, \orgname{Russian Academy of Sciences}, \orgaddress{\state{Nizhniy Arkhyz}, \country{Russia}}}

\address[3]{\orgdiv{Institute of Physics}, \orgname{Kazan Federal University}, \orgaddress{\state{Kazan}, \country{Russia}}}

\address[4]{\orgname{Institute of Astronomy, Czech Academy of Sciences}, \orgaddress{\state{Ond\v{r}ejov}, \country{Russia}}}

\corres{*Sergey Karpov, Institute of Physics of Czech Academy of Sciences, Na Slovance 1999/2, 182 21 Praha 8, Czech Republic \email{karpov.sv@gmail.com}}

\presentaddress{Sergey Karpov, Institute of Physics of Czech Academy of Sciences, Na Slovance 1999/2, 182 21 Praha 8, Czech Republic}

\abstract{
  To assess the potential of a small telescopes for a high temporal resolution astrophysics, we observed the field of FRB~121102 repeating source of fast radio bursts on a 50-cm D50 telescope of Ond\v{r}ejov observatory, equipped with a fast frame rate EMCCD detector. In a three nights of observations we did not detect any optical flares from the source, which allows us to place an upper limit of 10 mJy for a brightness of possible fast optical events on a time scale of 10 ms. We also characterize the apparent brightness stability of a field stars on the same time scale in order to investigate the potential of such telescopes for detecting faint optical variability on a sub-second time scales.}

\keywords{instrumentation: detectors; methods: statistical;  techniques: photometric}

\jnlcitation{\cname{%
\author{S. Karpov},
\author{M. Jelinek},
\author{J. \v{S}trobl}} (\cyear{2019}),
\ctitle{Fast optical bursts from FRB121102}, \cvol{2019;00:1--6}.}


\maketitle


\section{Introduction}\label{sec_intro}

 Fast radio bursts (FRBs) are bright and very short flashes of radio waves of yet unknown physical origin, first discovered more than 10 years ago \citep{lorimer_2007}.  They typically last less than a few milliseconds, and display a large dispersion measure, suggesting compact size and an extra-galactic origin. However, to date no counterparts have ever be detected in other wavebands, and no energetic transient events (gamma-ray bursts, supernovae, active galactic nuclei) have been associated with FRBs.

 The discovery of a repeating source of fast radio bursts, FRB~121102 \citep{spitler_2016},  opened unprecedented possibilities for an accurate determination of its position, and the source has been localized to a faint dwarf galaxy at a $z=0.19$ \citep{chatterjee_2017,tendulkar_2017}.

 To date, the only attempt for an optical monitoring of FRB~121102 position with high temporal resolution  and synchronous with radio observations was  published by \citet{hardy_2017}, where authors did not detect any flashes brighter than 1.2 mJy in 70.7 ms frames coincident with the times of radio bursts from the source. Therefore, we decided to observe the position of FRB~121102 synchronous with a dedicated radio monitoring run on a 100-m Effelsberg radio telescope \citep{shearer_private}. Unfortunately, there were no radio detections of a bursts from the object during that run \citep{shearer_private}, so we are unable to characterize the brightness of their simultaneous optical components. However, we may still place an upper limit on a rapid optical activity of the source during its radio quiet phase, as well as to characterize the performance of a 50-cm telescope with fast EMCCD camera for such a task.

 The paper is organized as follows. Section~\ref{sec_observations} describes the observations and data reduction, Section~\ref{sec_flashes} gives an overview of our search for a rapid optical flashes on individual frames. Then, in Section~\ref{sec_stability} we assess the stability of photometric measurements of a field stars on a time scale of 10-20 ms, and finally the Section~\ref{sec_conclusions} gives a brief conclusions.

\begin{center}
\begin{table*}[t]%
\caption{Summary of observations of a region around FRB~121102 on D50 telescope of Ond\v{r}ejov observatory, Czech Republic, in Sep 2017.\label{tab_observations}}%
\centering
\begin{tabular*}{500pt}{@{\extracolsep\fill}lccccccc@{\extracolsep\fill}}
\toprule
\textbf{Start time, UT} & \textbf{Filter}  & \textbf{Window}  & \textbf{Binning}  & \textbf{Exposure, s.}  & \textbf{FPS}  & \textbf{Duration, s.} & \textbf{Nframes} \\
\midrule
2017-09-27 00:21:06 & I & 512 & 2x2 & 0.02 & 47 & 9811 & 447903 \\
2017-09-27 03:05:57 & N & 512 & 2x2 & 0.02 & 47 & 2855 & 134132 \\
2017-09-28 01:50:23 & N & 512 & 2x2 & 0.02 & 47 & 7.5 & 355 \\
2017-09-28 01:55:54 & N & 256 & 2x2 & 0.01 & 86 & 7465 & 643522 \\
2017-09-28 23:33:22 & N & 256 & 2x2 & 0.01 & 86 & 15934 & 1330690 \\

\bottomrule
\end{tabular*}
\end{table*}
\end{center}

\section{Observations and data reduction}\label{sec_observations}

\begin{figure*}[!t]
  \centerline{
    \resizebox*{2.0\columnwidth}{!}{\includegraphics[angle=0]{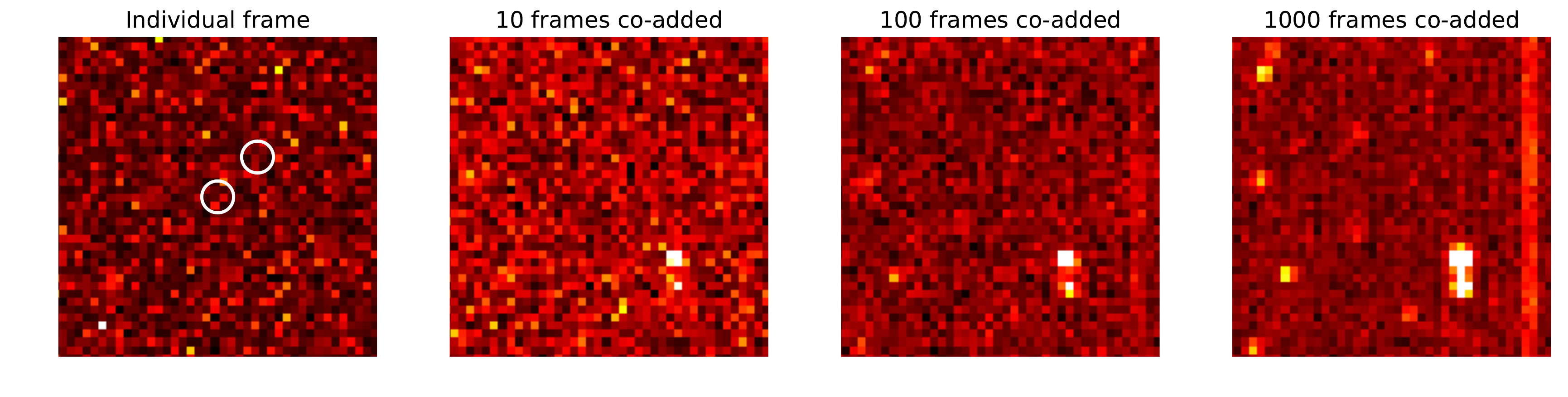}}
  }
  \caption{1.5$'$x1.5$'$ surroundings of a FRB~121102 position, as seen on individual frame (leftmost) or on a co-addition of 10, 100 and 1000 consecutive frames. Also, the  apertures used for extraction of object and background fluxes are shown.
    \label{fig_coadds}}
\end{figure*}

\begin{figure}[t]
  \centerline{
    \resizebox*{1.0\columnwidth}{!}{\includegraphics[angle=0]{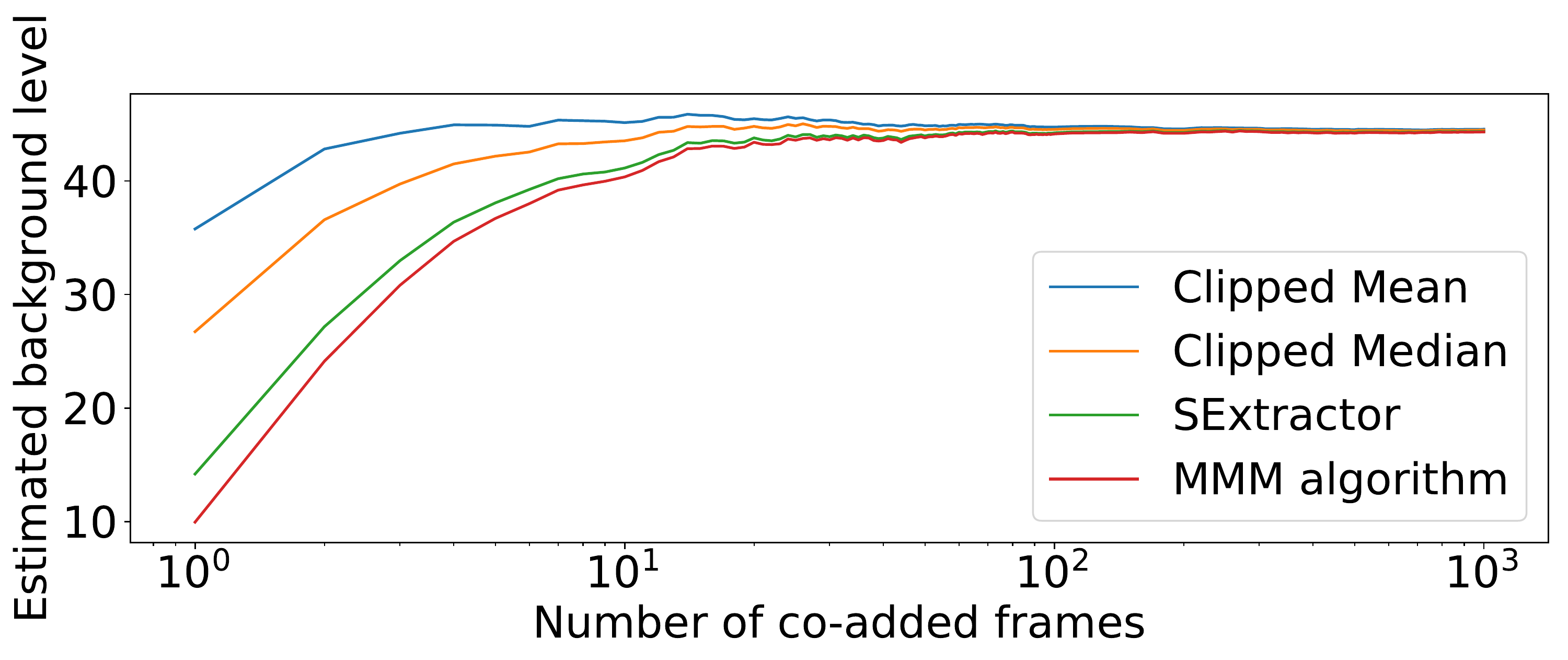}}
  }
  \caption{Background level estimation as a function of number of co-added frames for various algorithms. Due to significantly non-normal distribution of prevalent sources of noise, all of them are significantly biased on an individual frame.
    \label{fig_bg_est}}
\end{figure}

We observed the field of FRB121102 for a three nights in Sep 2017 with a D50 telescope \citep{d50} of Astronomical Institute of Czech Academy of Sciences, located in Ond\v{r}ejov observatory, Czech Republic. The telescope is a 500 mm f/4.5 Newtonian, equipped with standard B, V, R, I and N (Clear) filters, installed on a robotic mount and driven by {\sc RTS2} robotic observatory control software \citep{rts2}. During our observations, the telescope was equipped with Andor iXon DU-888 EMCCD, operated in an electron-multiplying regime with full amplification (300x). The detector has 1024x1024 13$\mu$m pixels, which gives 1.18$''$/pixel scale and a 20$'$x20$'$ field of view. To improve the temporal resolution, we used 2x2 binning for the whole run, and either full-frame or central half-frame readouts (which gives either 47 or 86 FPS frame rates).

The architecture of camera drivers in {\sc RTS2} does not support taking continuous sequences of images and/or working with high frame rate sensors. Thus, during the observations, the camera was controlled by a dedicated {\sc FAST} data acquisition software \citep{fast}, which is specifically designed for such scenarios. Using it, we acquired more than 2.5 millions of science frames in various regimes, summarized in Table~\ref{tab_observations}.

All these frames were dark subtracted using masterdarks constructed from a long series of ``dark'' frames acquired with the same gain, EM gain and exposure settings of the camera as scientific data, and then flatfield corrected using evening sky flats acquired in a normal manner during standard telescope operation. Every 30 consecutive frames were co-added and astrometrically calibrated using local {\sc Astrometry.Net} \citep{astrometry.net} installation to account for a tracking instabilities of the telescope mount and atmospheric variations.

Due to a significantly non-Gaussian nature of a noise that is dominant on short exposures (read-out and electron multiplication noises, clock-induced charges), an accurate background estimation on a single frame (see leftmost panel of Figure~\ref{fig_coadds} for an example of an individual frame) is  problematic (see Figure~\ref{fig_bg_est} for a dependence of several standard background estimators on a number of co-added frames).
Therefore, we derived the background map by employing a SExtractor-like mode estimator, as implemented in a {\sc SEP} \citep{sep} Python package, based on original {\sc SExtractor} code by \citet{sextractor}, on a 30 frames long co-added images, and using this estimation for analyzing individual frames.

To derive the photometric zero point, we used the stars from APASS DR9 \citep{apass} catalogue and the following photometric equations derived from analyzing long co-added images:
$$
\rm{Instr} = I - 0.12\cdot\left(B-V\right)\ \ \ \ \ \mbox{for I filter}
$$
$$
\rm{Instr} = V - 0.50\cdot\left(B-V\right)\ \ \ \ \ \mbox{for N filter}
$$

\section{Search for optical flashes}\label{sec_flashes}

\begin{figure}[t]
  \centerline{
    \resizebox*{1.0\columnwidth}{!}{\includegraphics[angle=0]{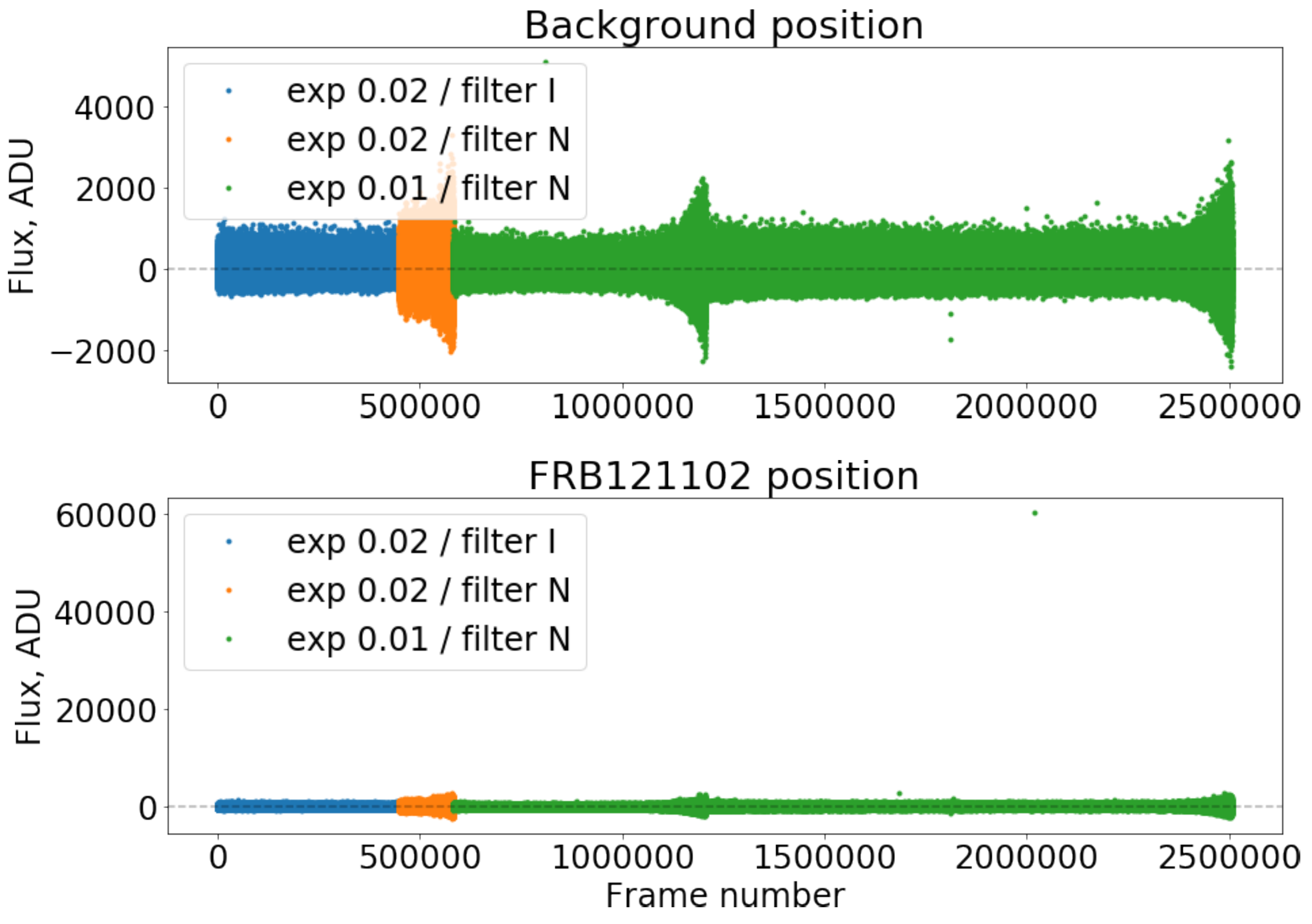}}
  }
  \caption{Fluxes at the position of FRB~121102 and at a comparison background position, measured in a circular apertures with 2 pixels radius on all frames acquired in three nights of observations. Mean values of both curves are zero. Time-dependent variance is due to the background sky level increase towards the morning. Singular outliers are due to either read-out defects, cosmic ray hits or other non-astrophysical sources.
    \label{fig_fluxes}}
\end{figure}

\begin{figure}[t]
  \centerline{
    \resizebox*{1.0\columnwidth}{!}{\includegraphics[angle=0]{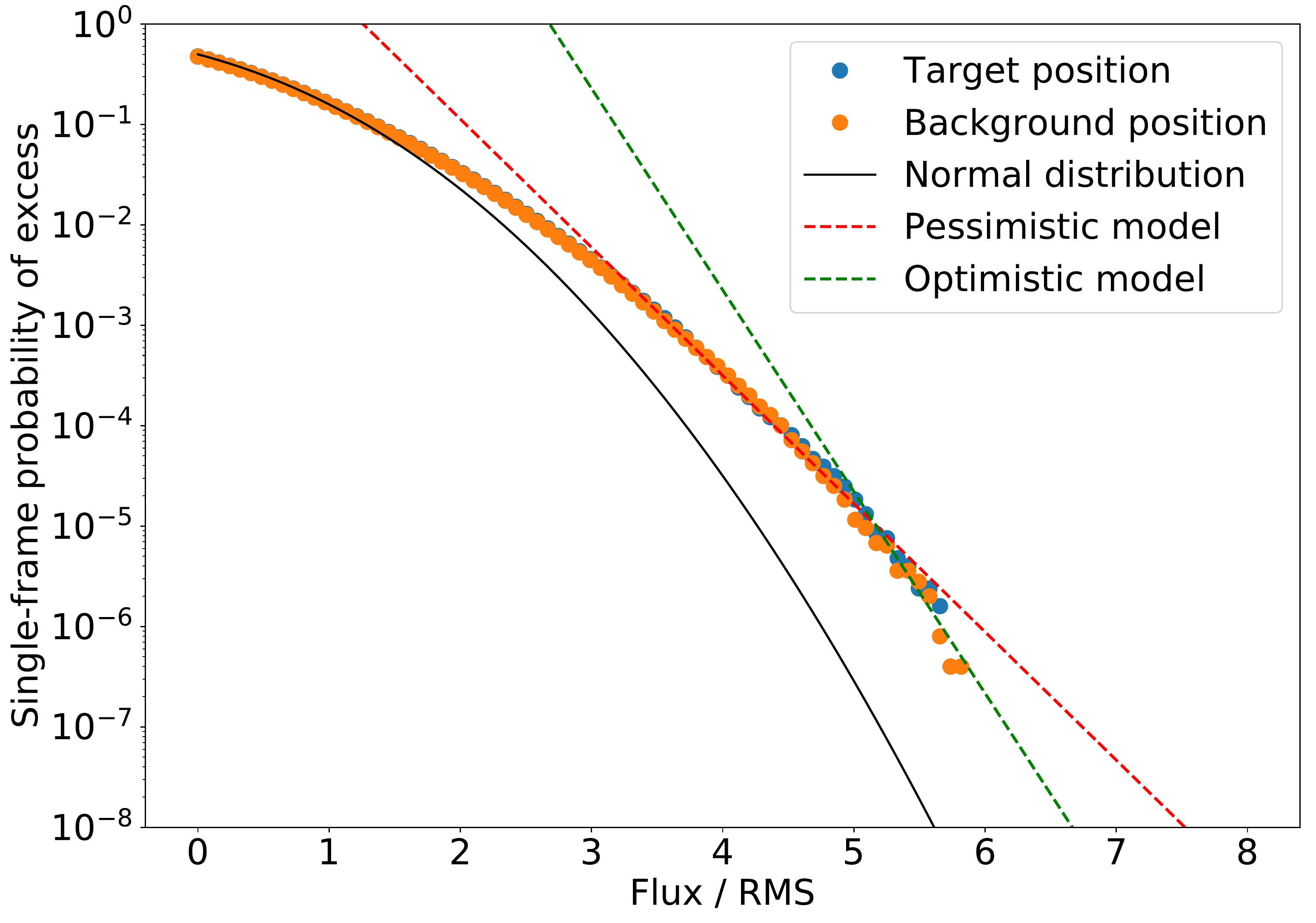}}
  }
  \caption{Survival functions (probabilities of surpassing a given value) for a RMS-normalized fluxes at the position of FRB~121102 and a comparison background position. Also shown is the same function for a normal distribution. Two extrapolant models are shown what should accommodate for a slope of possible rightward extension of survival functions.
    \label{fig_survival}}
\end{figure}

In order to detect possible optical flashes from FRB~121102, we performed an aperture photometry with 2 pixels radius at a target position of FRB121102 and a background position close to it (see leftmost panel of Figure~\ref{fig_coadds} for a positions of object and background apertures, and Figure~\ref{fig_fluxes} for a lightcurves inside these apertures), and compared their statistical properties looking for any outliers. All of a few statistically significant deviations of target flux were visually inspected and found to be non-astrophysical, related to either detector read-out anomalies, cosmic ray hits or meteors.

To define the upper limit for a single-frame optical flashes detectable as a significant outliers in our observations,
we normalized both fluxes to the running estimations of standard deviation,
and then computed a survival functions for them, shown in Figure~\ref{fig_survival}. Both are practically identical, and both display larger amount of high-amplitude outliers than expected for a normal distribution, as expected for a large amplification regimes of EMCCDs.
We then constructed two power-law extrapolants for an outward slope of survival functions, ``optimistic'' (steeper) and ``pessimistic'' (flatter) ones, which should constrain their possible behaviour. Finally, we used these extrapolants to get the flux levels corresponding to a $10^{-7}$ probability (effectively ``5$\sigma$'') for a number of trials corresponding to the one of acquired frames (see Table~\ref{tab_observations}). The results for the upper limits for a brightness of possible events on the time scale of single exposure for different regimes used in our observations are listed in Table~\ref{tab_limits}.

\begin{center}
\begin{table}[t]%
\centering
\caption{Upper limits for a brightness of possible optical flashes at the position of FRB~121102 on time scale of a single exposure.\label{tab_limits}}%
\tabcolsep=0pt%
\begin{tabular*}{20pc}{@{\extracolsep\fill}ccc@{\extracolsep\fill}}
\toprule
\textbf{Regime} & \textbf{``Optimistic'' limit}  & \textbf{``Pessimistic'' limit} \\
\midrule
0.02 s / filter I &	12.9m = 18.2 mJy &	12.8m = 20 mJy \\
0.02 s / filter N &	14.0m = 9.5 mJy &	13.8m = 11.4 mJy  \\
0.01 s / filter N &	13.8m = 11.4 mJy &	13.6m = 13.7 mJy \\
\bottomrule
\end{tabular*}
\end{table}
\end{center}

\section{Photometric stability}\label{sec_stability}

\begin{figure}[t]
  \centerline{
    \resizebox*{1.0\columnwidth}{!}{\includegraphics[angle=0]{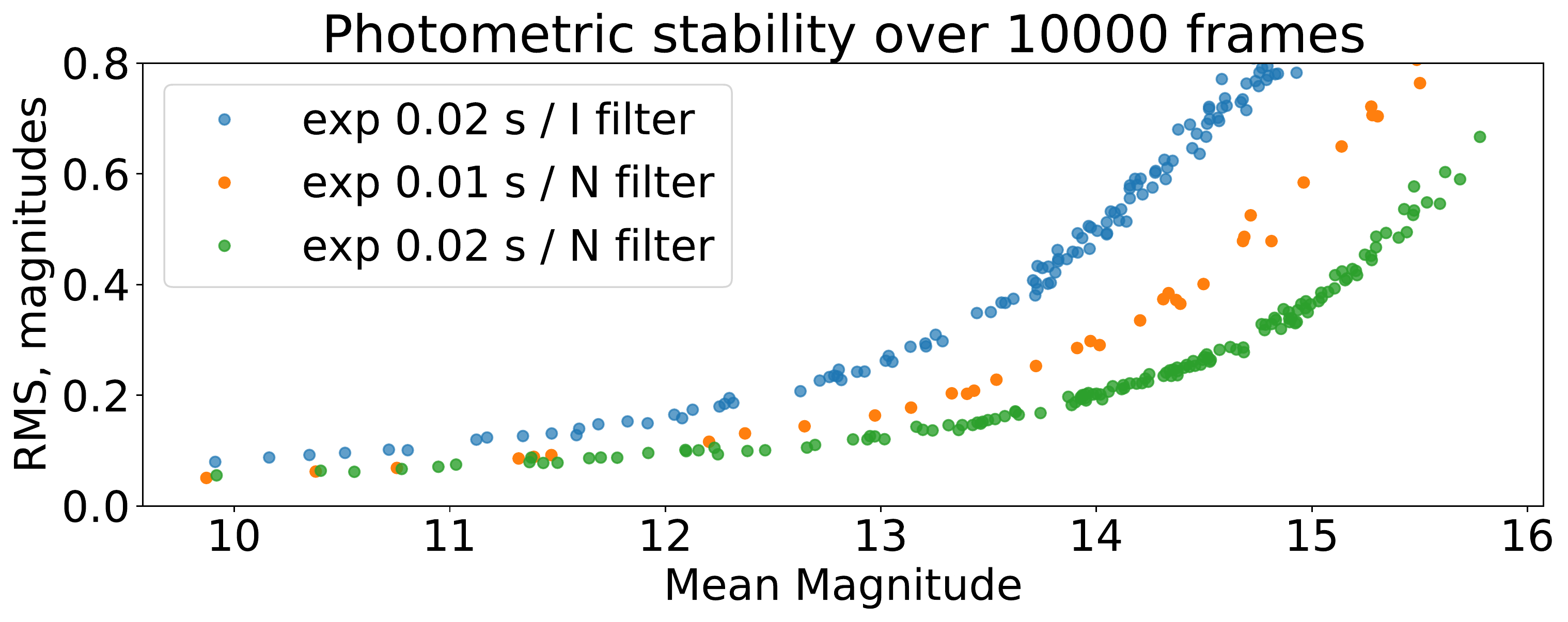}}
  }
  \caption{Scatter versus magnitude plot for a photometric measurements of stars over a long sequence of consecutive frames for different regimes we used in observations. The achievable precision is 0.05-0.1$^{\rm{m}}$ for brighter objects.
    \label{fig_mag_rms}}
\end{figure}

\begin{figure}[t]
  \centerline{
    \resizebox*{1.0\columnwidth}{!}{\includegraphics[angle=0]{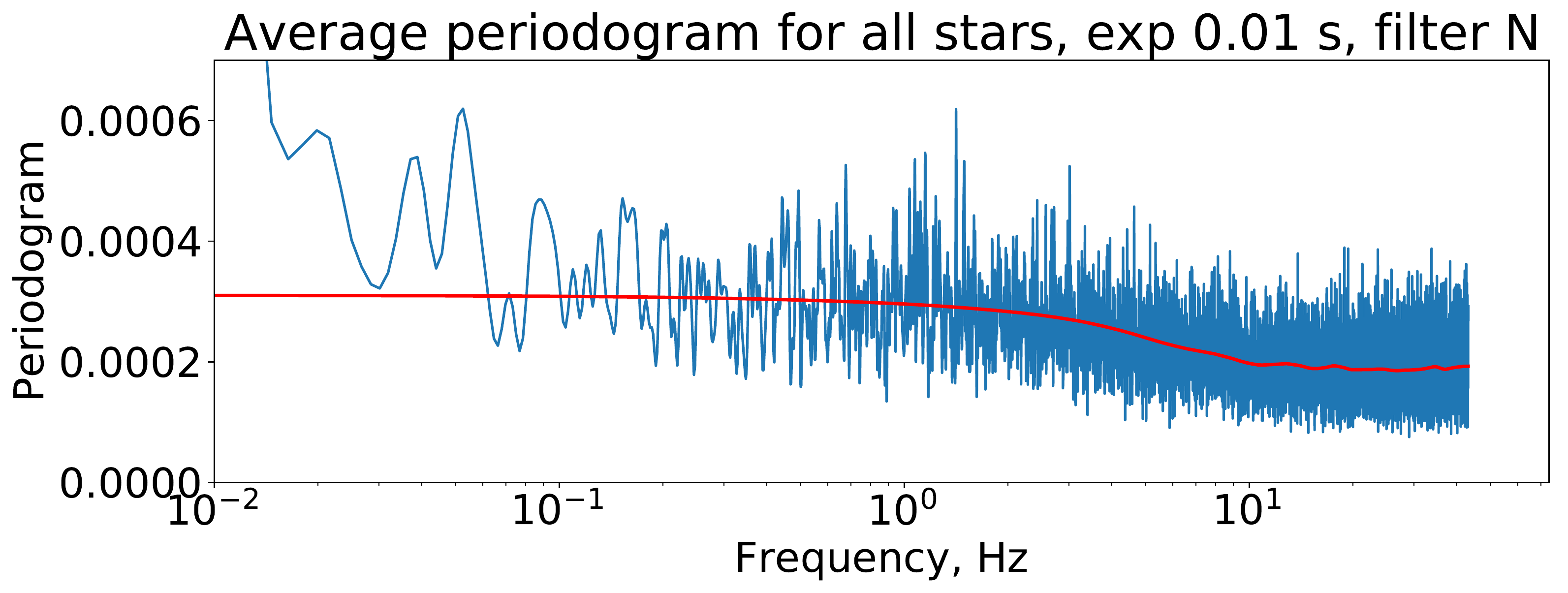}}
  }
  \caption{Average periodogram for 21 light curves acquired with N filter and 0.01 s exposures.
    \label{fig_periodogram}}
\end{figure}

In order to assess the limits of photometric stability achievable in high temporal resolution observations with D50 telescope, we performed a ``forced'' aperture photometry on a continuous sequences of individual frames, with aperture positions fixed to the ones measured on a co-addition of large number of frames. Background level was also estimated on a co-added frames in a way described in Section~\ref{sec_observations}. Zero points were derived on every individual frame by fitting the ensemble brightness of the stars with catalogue values corrected for a photometric equation of a specific filter.
We then computed mean values and standard deviations of every light curve, and constructed the scatter-magnitude diagram (see Figure~\ref{fig_mag_rms}) in order to check how precise are the individual measures and what one may expect from a high temporal resolution observations on a such telescope. About 5\% accuracy is achievable for brighter objects on a time scale of 10-20 ms.

Periodogram analysis (averaging of individual periodograms of every extracted light curve, see Figure~\ref{fig_periodogram}) displays a slight power increase below several Hz, which may be attributed to either atmospheric instabilities or a slow drifts of an electron-multiplying gain of the camera over time, and needs a more detailed investigation. Probably, proper accounting for it may further improve the achievable precision levels of photometric observations.

\section{Conclusions}\label{sec_conclusions}

In our three nights of observations of FRB~121102 position with 50-cm D50 telescope in Sep 2017, resulted in acquisition of more than 2.5 millions of individual frames, we did not detect any significant optical flashes at the position of the source. The non-detection allows us to define an upper level of the possible brightness of flashes on the level of 10 to 20 mJy in 10-20 ms, depending on the filter used. As there was no radio detection of activity from the source during our observations \citep{shearer_private}, we can't directly compare our results with the ones of \citet{hardy_2017}, however, their single-frame upper limits of 1.2 mJy in 70.7 ms exposures makes an actual performance of D50 telescope quite similar for bursts shorter than out 10ms exposure, and the more so -- considering the number-of-trials corrections we had to use to derive the limits due to absence of ``selected'' time positions.

Therefore we may conclude that even such small, 50-cm telescopes, when equipped with fast and low-noise detectors, may be quite useful tools for a monitoring of a rare astrophysical events with known localizations, like repeating FRB sources, flaring stars, X-ray binaries etc.


\section*{Acknowledgments}

This work was supported by \fundingAgency{European Structural and Investment Fund} and the \fundingAgency{Czech Ministry of Education, Youth and Sports} (Project CoGraDS -- \fundingNumber{CZ.02.1.01/0.0/0.0/15 003/0000437}). The work is performed in the framework of the Russian Government Program of Competitive Growth of the Kazan Federal University.

\subsection*{Author contributions}

S.K. participated in observations and performed all data analysis. M.J. and J.\v{S}. maintained the telescope and equipment used in observations, and participated in observations.

\subsection*{Financial disclosure}

None reported.

\subsection*{Conflict of interest}

The authors declare no potential conflict of interests.

\bibliography{frb}%

\begin{thebibliography}{}

\bibitem [\protect \citeauthoryear {%
{Barbary}%
}{%
{Barbary}%
}{%
{\protect \APACyear {2018}}%
}]{%
sep}
\APACinsertmetastar {%
sep}%
\begin{APACrefauthors}%
{Barbary}, K.%
\end{APACrefauthors}%
\unskip\
\newblock
\APACrefYearMonthDay{2018}{Nov}{},
\newblock
\APACrefbtitle {{SEP: Source Extraction and Photometry}.} {{SEP: Source
  Extraction and Photometry}.}
\PrintBackRefs{\CurrentBib}

\bibitem [\protect \citeauthoryear {%
{Bertin}%
\ \BBA {} {Arnouts}%
}{%
{Bertin}%
\ \BBA {} {Arnouts}%
}{%
{\protect \APACyear {1996}}%
}]{%
sextractor}
\APACinsertmetastar {%
sextractor}%
\begin{APACrefauthors}%
{Bertin}, E.%
\BCBT {}\ \BBA {} {Arnouts}, S.%
\end{APACrefauthors}%
\unskip\
\newblock
\APACrefYearMonthDay{1996}{{\APACmonth{06}}}{},
\newblock
\unskip
\newblock
\APACjournalVolNumPages{\aaps}{117}{}{393-404}.
\newblock
\begin{APACrefDOI} \doi{10.1051/aas:1996164} \end{APACrefDOI}
\PrintBackRefs{\CurrentBib}

\bibitem [\protect \citeauthoryear {%
{Chatterjee}%
\ \protect \BOthers {.}}{%
{Chatterjee}%
\ \protect \BOthers {.}}{%
{\protect \APACyear {2017}}%
}]{%
chatterjee_2017}
\APACinsertmetastar {%
chatterjee_2017}%
\begin{APACrefauthors}%
{Chatterjee}, S.%
, {Law}, C\BPBI J.%
, {Wharton}, R\BPBI S.%
\ et al.\end{APACrefauthors}%
\unskip\
\newblock
\APACrefYearMonthDay{2017}{Jan}{},
\newblock
\unskip
\newblock
\APACjournalVolNumPages{\nat}{541}{7635}{58-61}.
\newblock
\begin{APACrefDOI} \doi{10.1038/nature20797} \end{APACrefDOI}
\PrintBackRefs{\CurrentBib}

\bibitem [\protect \citeauthoryear {%
{Hardy}%
\ \protect \BOthers {.}}{%
{Hardy}%
\ \protect \BOthers {.}}{%
{\protect \APACyear {2017}}%
}]{%
hardy_2017}
\APACinsertmetastar {%
hardy_2017}%
\begin{APACrefauthors}%
{Hardy}, L\BPBI K.%
, {Dhillon}, V\BPBI S.%
, {Spitler}, L\BPBI G.%
\ et al.\end{APACrefauthors}%
\unskip\
\newblock
\APACrefYearMonthDay{2017}{Dec}{},
\newblock
\unskip
\newblock
\APACjournalVolNumPages{\mnras}{472}{3}{2800-2807}.
\newblock
\begin{APACrefDOI} \doi{10.1093/mnras/stx2153} \end{APACrefDOI}
\PrintBackRefs{\CurrentBib}

\bibitem [\protect \citeauthoryear {%
{Henden}%
, {Levine}%
, {Terrell}%
\BCBL {}\ \BBA {} {Welch}%
}{%
{Henden}%
\ \protect \BOthers {.}}{%
{\protect \APACyear {2015}}%
}]{%
apass}
\APACinsertmetastar {%
apass}%
\begin{APACrefauthors}%
{Henden}, A\BPBI A.%
, {Levine}, S.%
, {Terrell}, D.%
\BCBL {}\ \BBA {} {Welch}, D\BPBI L.%
\end{APACrefauthors}%
\unskip\
\newblock
\APACrefYearMonthDay{2015}{Jan}{},
\newblock
{\BBOQ}\APACrefatitle {{APASS - The Latest Data Release}} {{APASS - The Latest
  Data Release}}.{\BBCQ}
\newblock
\BIn{} \APACrefbtitle {American Astronomical Society Meeting Abstracts \#225}
  {American Astronomical Society Meeting Abstracts \#225}\ \BVOL~225,
  \BPG~336.16.
\PrintBackRefs{\CurrentBib}

\bibitem [\protect \citeauthoryear {%
Karpov%
}{%
Karpov%
}{%
{\protect \APACyear {2018}}%
}]{%
fast}
\APACinsertmetastar {%
fast}%
\begin{APACrefauthors}%
Karpov, S.%
\end{APACrefauthors}%
\unskip\
\newblock
\APACrefYearMonthDay{2018}{}{},
\newblock
\APACrefbtitle {FAST data acquisition software.} {FAST data acquisition
  software.},
\newblock
\APAChowpublished {\url{https://github.com/karpov-sv/fast}}.
\newblock
\APACaddressPublisher{}{GitHub}.
\PrintBackRefs{\CurrentBib}

\bibitem [\protect \citeauthoryear {%
{Kub{\'a}nek}%
\ \protect \BOthers {.}}{%
{Kub{\'a}nek}%
\ \protect \BOthers {.}}{%
{\protect \APACyear {2004}}%
}]{%
rts2}
\APACinsertmetastar {%
rts2}%
\begin{APACrefauthors}%
{Kub{\'a}nek}, P.%
, {Jel{\'{\i}}nek}, M.%
, {Nekola}, M.%
\ et al.\end{APACrefauthors}%
\unskip\
\newblock
\APACrefYearMonthDay{2004}{{\APACmonth{09}}}{},
\newblock
{\BBOQ}\APACrefatitle {{RTS2 - Remote Telescope System, 2$^{nd}$ Version}}
  {{RTS2 - Remote Telescope System, 2$^{nd}$ Version}}.{\BBCQ}
\newblock
\BIn{} E.~{Fenimore}\ \BBA {} M.~{Galassi}\ (\BEDS), \APACrefbtitle {Gamma-Ray
  Bursts: 30 Years of Discovery} {Gamma-Ray Bursts: 30 Years of Discovery}\
  \BVOL~727, \BPG~753-756.
\newblock
\begin{APACrefDOI} \doi{10.1063/1.1810951} \end{APACrefDOI}
\PrintBackRefs{\CurrentBib}

\bibitem [\protect \citeauthoryear {%
{Lang}%
, {Hogg}%
, {Mierle}%
, {Blanton}%
\BCBL {}\ \BBA {} {Roweis}%
}{%
{Lang}%
\ \protect \BOthers {.}}{%
{\protect \APACyear {2010}}%
}]{%
astrometry.net}
\APACinsertmetastar {%
astrometry.net}%
\begin{APACrefauthors}%
{Lang}, D.%
, {Hogg}, D\BPBI W.%
, {Mierle}, K.%
, {Blanton}, M.%
\BCBL {}\ \BBA {} {Roweis}, S.%
\end{APACrefauthors}%
\unskip\
\newblock
\APACrefYearMonthDay{2010}{May}{},
\newblock
\unskip
\newblock
\APACjournalVolNumPages{\aj}{139}{5}{1782-1800}.
\newblock
\begin{APACrefDOI} \doi{10.1088/0004-6256/139/5/1782} \end{APACrefDOI}
\PrintBackRefs{\CurrentBib}

\bibitem [\protect \citeauthoryear {%
{Lorimer}%
, {Bailes}%
, {McLaughlin}%
, {Narkevic}%
\BCBL {}\ \BBA {} {Crawford}%
}{%
{Lorimer}%
\ \protect \BOthers {.}}{%
{\protect \APACyear {2007}}%
}]{%
lorimer_2007}
\APACinsertmetastar {%
lorimer_2007}%
\begin{APACrefauthors}%
{Lorimer}, D\BPBI R.%
, {Bailes}, M.%
, {McLaughlin}, M\BPBI A.%
, {Narkevic}, D\BPBI J.%
\BCBL {}\ \BBA {} {Crawford}, F.%
\end{APACrefauthors}%
\unskip\
\newblock
\APACrefYearMonthDay{2007}{Nov}{},
\newblock
\unskip
\newblock
\APACjournalVolNumPages{Science}{318}{5851}{777}.
\newblock
\begin{APACrefDOI} \doi{10.1126/science.1147532} \end{APACrefDOI}
\PrintBackRefs{\CurrentBib}

\bibitem [\protect \citeauthoryear {%
{Nekola}%
\ \protect \BOthers {.}}{%
{Nekola}%
\ \protect \BOthers {.}}{%
{\protect \APACyear {2010}}%
}]{%
d50}
\APACinsertmetastar {%
d50}%
\begin{APACrefauthors}%
{Nekola}, M.%
, {Hudec}, R.%
, {Jel{\'\i}nek}, M.%
\ et al.\end{APACrefauthors}%
\unskip\
\newblock
\APACrefYearMonthDay{2010}{Aug}{},
\newblock
\unskip
\newblock
\APACjournalVolNumPages{Experimental Astronomy}{28}{1}{79-85}.
\newblock
\begin{APACrefDOI} \doi{10.1007/s10686-010-9190-5} \end{APACrefDOI}
\PrintBackRefs{\CurrentBib}

\bibitem [\protect \citeauthoryear {%
Shearer%
}{%
Shearer%
}{%
{\protect \APACyear {2017}}%
}]{%
shearer_private}
\APACinsertmetastar {%
shearer_private}%
\begin{APACrefauthors}%
Shearer, A.%
\end{APACrefauthors}%
\unskip\
\newblock
\APACrefYearMonthDay{2017}{}{},
\newblock
\APAChowpublished {private communication}.
\PrintBackRefs{\CurrentBib}

\bibitem [\protect \citeauthoryear {%
{Spitler}%
\ \protect \BOthers {.}}{%
{Spitler}%
\ \protect \BOthers {.}}{%
{\protect \APACyear {2016}}%
}]{%
spitler_2016}
\APACinsertmetastar {%
spitler_2016}%
\begin{APACrefauthors}%
{Spitler}, L\BPBI G.%
, {Scholz}, P.%
, {Hessels}, J\BPBI W\BPBI T.%
\ et al.\end{APACrefauthors}%
\unskip\
\newblock
\APACrefYearMonthDay{2016}{Mar}{},
\newblock
\unskip
\newblock
\APACjournalVolNumPages{\nat}{531}{7593}{202-205}.
\newblock
\begin{APACrefDOI} \doi{10.1038/nature17168} \end{APACrefDOI}
\PrintBackRefs{\CurrentBib}

\bibitem [\protect \citeauthoryear {%
{Tendulkar}%
\ \protect \BOthers {.}}{%
{Tendulkar}%
\ \protect \BOthers {.}}{%
{\protect \APACyear {2017}}%
}]{%
tendulkar_2017}
\APACinsertmetastar {%
tendulkar_2017}%
\begin{APACrefauthors}%
{Tendulkar}, S\BPBI P.%
, {Bassa}, C\BPBI G.%
, {Cordes}, J\BPBI M.%
\ et al.\end{APACrefauthors}%
\unskip\
\newblock
\APACrefYearMonthDay{2017}{Jan}{},
\newblock
\unskip
\newblock
\APACjournalVolNumPages{\apj}{834}{2}{L7}.
\newblock
\begin{APACrefDOI} \doi{10.3847/2041-8213/834/2/L7} \end{APACrefDOI}
\PrintBackRefs{\CurrentBib}

\end{thebibliography}

\section*{Author Biography}

\begin{biography}
  {\includegraphics[width=70pt,height=70pt,draft]{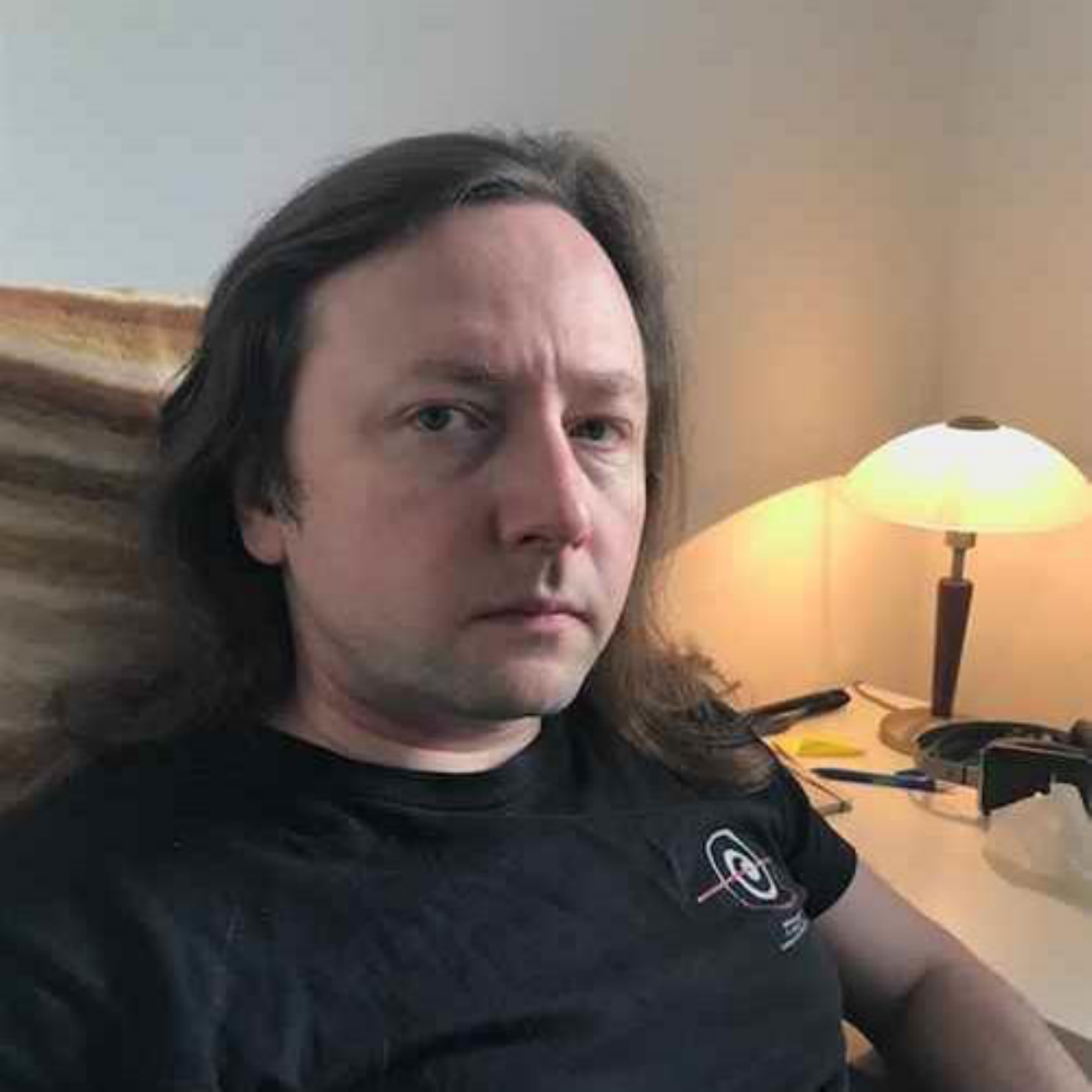}}
  {\textbf{Karpov, Sergey.}
    Sergey Karpov finished Moscow State University in 2003, and got his PhD in astrophysics in Special Astrophysical Observatory, Russia, in 2007. His scientific interests include high temporal resolution astrophysics, time-domain sky surveys, astronomical data processing pipelines and transient detection algorighms. He is currently working on development of automated data analysis pipelines for several time domain sky surveys, as well as on various aspects of testing and characterization of optical detectors used in astronomy.}

\end{biography}

\end{document}